\documentclass{elsart}               

\usepackage{graphicx}          
\usepackage[isolatin]{inputenc}
\usepackage{dcolumn}
\usepackage{bm}
\begin{document}

\begin{frontmatter}

\title{\emph{Q}\ value of the $^{100}$Mo Double-Beta Decay} 

\author{S. Rahaman\corauthref{cor1}}
\ead{saidur.rahaman@phys.jyu.fi},
\author{V.-V. Elomaa},
\author{T. Eronen},
\author{J. Hakala},
\author{A. Jokinen},
\author{J. Julin},
\author{A. Kankainen},
\author{A. Saastamoinen},
 \author{J. Suhonen},
 \author{C. Weber} and
 \author{J. Äystö}
\corauth[cor1]{Corresponding author}
\address{Department of Physics, P.O. Box 35 (YFL), FIN-40014 University of Jyväskylä, Finland}  


       


\begin{abstract}                          
Penning trap measurements using mixed beams of $^{100}$Mo - $^{100}$Ru and $^{76}$Ge - $^{76}$Se have been utilized to determine the double-beta decay 
\emph{Q}-values of $^{100}$Mo and $^{76}$Ge with uncertainties less than 200 eV. The value for $^{76}$Ge, 2039.04(16) keV is in agreement with the 
published SMILETRAP value. The new value for $^{100}$Mo, 3034.40(17) keV is 30 times more precise than the previous literature value, suffcient for 
the ongoing neutrinoless double-beta decay searches in $^{100}$Mo. Moreover, the precise \emph{Q}-value is used to calculate the phase-space integrals 
and the experimental nuclear matrix element of double-beta decay.
\end{abstract}

\end{frontmatter}

\section{Introduction}
Neutrinos are one of the least understood fundamental particles. For half a century physicists thought that neutrinos, like photons, had no mass. But 
recent data from the neutrino oscillation experiments at SuperKamiokande \cite{sk}, SNO \cite{sn}, and KamLAND \cite{kl} overturned this view and 
confirmed that the neutrinos are massive particles. However, oscillation experiments can yield only the differences in the square of neutrino masses, 
therefore, no absolute mass values can be determined. In addition, another question remains concerning the fundamental character of neutrinos, whether 
being Dirac or Majorana particles. Neutrinoless double-beta decay ($0\nu\beta\beta$-decay) is a process which can address both issues raised above. 
This decay process is forbidden according to the Standard Model of Particle Physics since it violates the lepton-number conservation and is only 
allowed if neutrinos are massive Majorana particles. 

The Heidelberg-Moscow Collaboration \cite{LB99} has claimed the observation of the $0\nu\beta\beta$-decay using high-sensitivity \cite{HV011,CE02} 
$^{76}$Ge semiconductor detectors \cite{HV01}. $^{100}$Mo is another suitable nucleus to study the $0\nu\beta\beta$-decay for the following reasons. 
$^{100}$Mo is rather easy to produce as enriched material and its $2\nu\beta\beta$ matrix element is known \cite{HE91}. Furthermore, it has a high 
\emph{Q}-value and the $0\nu\beta\beta$-decay rate scales with $Q^{5}$. The MOON \cite{MN05} and NEMO-3 \cite{RA05} experiments aim to use $^{100}$Mo 
for the $0\nu\beta\beta$-decay search. The signal for a $0\nu\beta\beta$-decay in a detector measuring the total energy of both electrons would be a 
peak at the position of the \emph{Q}-value of the involved transition. Thus it is vital for the ongoing search of the $0\nu\beta\beta$-decay that the 
\emph{Q}-value is known to a fraction of the expected detector resolution. 

Additional motivation for the precise \emph{Q}-value measurement is to calculate the precise phase-space integral \emph{G} and the experimental 
nuclear matrix element of the two-neutrino double-deta decay \cite{JS98}. The half-life $T_{1/2}$ of the double-beta decay process is expressed as 
\begin{equation}
[T_{1/2}^{2\nu}]^{-1} = G^{2\nu}(M^{2\nu})^{2},
\end{equation}
for the $2\nu\beta\beta$-decay and 
\begin{equation}
[T_{1/2}^{0\nu}]^{-1} = G^{0\nu}(M^{0\nu})^{2} (\langle m_{\nu}\rangle / m_{e})^{2} ,
\end{equation}
for the $0\nu\beta\beta$-decay. Here \emph{M} is the nuclear matrix element between
the initial and final states of the decay, $\langle m_{\nu}\rangle$ is the effective neutrino mass  \cite{JS98} and $m_{e}$ is the electron rest mass. 
For the $2\nu\beta\beta$-decay case $T_{1/2}^{2\nu}$ can be derived from experiment and $G^{2\nu}$ calculated reliably. Hence using Eq.~(1) the 
experimental $M^{2\nu}$ is estimated. In the case of the $0\nu\beta\beta$-decay the experimental half-life is unknown and hence we give the half-life 
of  $^{100}$Mo as a function of the effective neutrino mass for a given set of matrix elements using the $G^{0\nu}$ value taken from this work.

 In this paper we present the precise double-beta decay \emph{Q}-value of $^{100}$Mo and report on the development of an off-line ion source for 
producing stable (or long-lived) mixed-ion beams for fundamental physics studies.
\section{Experimental method}
\begin{figure*}
\resizebox{0.99\textwidth}{!}{
\includegraphics{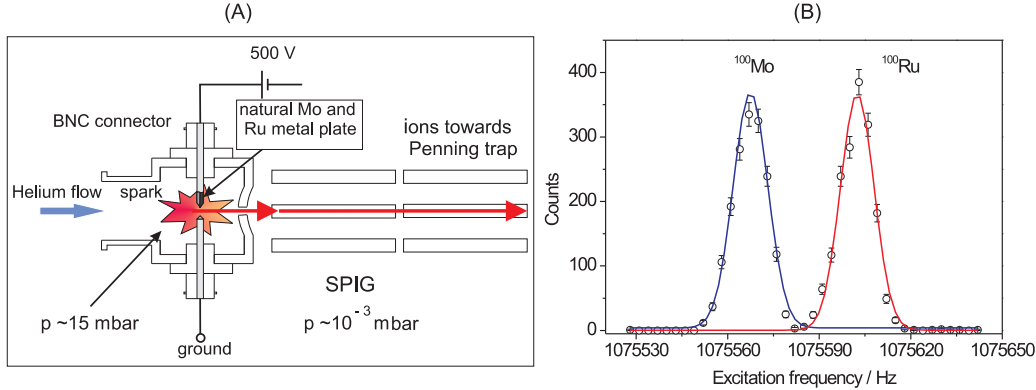}}
\caption{\label{fig:epsart}(\textbf{A}): Schematic drawing of an off-line ion source to produce a variety of stable and long-lived mixed 
singly-charged ion beams. (\textbf{B}): Mass spectrum of ions ejected from the purification trap after buffer gas cooling for mass region A = 100. A 
gaussian fit to the data yields a FWHM of about 11 Hz, which corresponds to a mass resolving power of about $1\times 10^{5}$.}
\end{figure*}
JYFLTRAP \cite{AJ06} is an ion trap experiment for high-precision mass measurements of radioactive ions \cite{UH06,AK06,SR07,SR07a,UH07}. JYFLTRAP has 
been used for direct \emph{$Q_{EC}$}-value measurements of the superallowed beta emitters \cite{TE06,TE06a,CW07} and recently for double-beta decay. 
Our \emph{Q}-value measurement is based on a comparison of the ion cyclotron frequency of a stored ion in a Penning trap \cite{GB86}. The cyclotron 
frequency $\nu_{c}$ of an ion is given by 

\begin{equation}
\nu_{c} = \frac{1}{2\pi}\frac{q}{m}B,
\end{equation}
where \emph{B} is the magnetic field, \emph{m} is the mass and \emph{q} the charge of the ion. The ions of interest are produced at the Ion Guide 
Isotope Separator On-line (IGISOL) facility \cite{JA01} which has an advantage that both mother and daughter nuclei can be produced in the same 
reaction. On the other hand, an off-line ion source was used to produce a mixed ion beam of long-lived (mother and daughter) nuclei. Thus it enabled 
the measurement of the cyclotron frequencies of the mother and daughter nucleus in a consecutive manner. Hence the \emph{Q}-value can be obtained by 
using the following formula:
\begin{equation}
Q = m_{m} - m_{d} = \left(\frac{\nu_{d}}{\nu_{m}}-1\right)m_{d},
\end{equation}
where $m_{m}$ and $m_{d}$ are the masses of the mother and daughter ions and $\frac{\nu_{d}}{\nu_{m}}$ is their cyclotron frequency ratio. The 
daughter nucleus was used as a reference ion and its mass excess value was obtained from ref. \cite{GA03}. 

In the case of the double-beta decay, $^{100}$Mo (mother nucleus) and $^{100}$Ru (daughter nucleus) ions were produced by using an off-line ion source 
shown in Fig.~1(\textbf{A}). The ion source consists of two electrodes and it is similar in size to the light ion guide used for indiced fusion 
reactions at the IGISOL facility \cite{JA01}. One of the electrodes was designed to hold the metal plates or powder of the enriched element of need 
and the other one was the ground electrode. To create ions a continuous spark was ignited by applying 500 V between the two electrodes at a 15 mbar 
helium pressure. In general this off-line ion source can be used to produce any stable mixed-ion beam. In this work it was used at JYFLTRAP to create 
$^{76}$Ge - $^{76}$Se and $^{100}$Mo - $^{100}$Ru pairs of singly-charged ions.

Ions were extracted from the off-line source by helium flow and guided by the sextupole ion guide (SPIG) into a differential pumping stage where they 
were accelerated to 30 keV and mass-separated with a 55$^{0}$ dipole magnet with a mass resolving power $M/\Delta M$ of $\sim$ 500. The ions were then 
transported to a radiofrequency quadrupole (RFQ) structure where they were cooled, accumulated and bunched \cite{AN03}. Finally, they were injected 
into the double Penning trap system. 

The JYFLTRAP Penning traps are placed inside the warm bore of a 7 T superconducting magnet and they are separated by a 2-mm channel. The first trap is 
called the purification Penning trap, where the mass-selective buffer-gas cooling technique is applied for further axial cooling and isobaric cleaning 
\cite{GS91}. The mass resolving power of the purification trap was on the order of $10^{5}$. Figure~1(\textbf{B}) displays a mass spectrum of ions 
ejected from the purification trap after the buffer gas cooling. The purification trap was operated to center selectively the mass region \emph{A} = 
100. The one of the selected ion samples was transported to the second trap called the precision trap.  

In addition, time-separated (Ramsey method) \cite{NFR90} dipolar excitation was applied in the precision trap to ensure a single ion species is 
selected \cite{TE07}. This was performed in the following way: at first the mass-selective reduced cyclotron frequency $\nu_{+}$ (removal frequency) 
was applied for one ion species as two time-separated fringes of 5 ms with a waiting time of 20 ms (5-20-5 ms). This increases the reduced cyclotron 
radius of the unwanted ions. At the end of the excitation the ions were sent back to the purification trap, thus the excited ions (\emph{i.e.} the 
unwanted ions) will not pass the 2-mm channel between the traps. This method allows us to clean even isomeric states that are only 200 keV higher than 
the ground state (in the case of $^{54}$Co) \cite{TE08}. In the purification trap, the buffer-gas cooling technique was re-applied and finally a pure 
ion sample was transported to the precision trap for the cyclotron frequency ($\nu_{c}$) measurement. 

At first, the cyclotron frequency was determined by employing the time-of-flight (TOF) technique \cite{MK95} with an excitation time of 100 ms in a 
single-fringe quadrupolar excitation scheme (conventional excitation scheme). Once the cyclotron frequency was determined, a quadrupolar Ramsey 
excitation was applied in order to determine the cyclotron frequency with higher precision \cite{SG07}. 

A typical TOF resonance is shown in Fig.~2 (top panel) for $^{100}$Ru ions with a conventional excitation scheme and an excitation time of 100 ms. The 
middle and bottom panels in Fig.~2  show the Ramsey resonances for $^{100}$Ru and $^{100}$Mo ions, respectively. A Ramsey excitation with two 
time-separated fringes of 50 ms and a waiting time of 400 ms (50-400-50 ms) was applied for these particular cases. 

\section{Analysis and Results}

The statistical standard deviation $\sigma (\nu_{c})$ of the cyclotron frequency  $\nu_{c}$ can be estimated as \cite{GB01}
\begin{equation}
\frac {\sigma (\nu_{c})}{\nu_{c}} = \frac{1}{\nu_{c}} \times \frac{K}{\sqrt{N}\cdot T_{ex}},
\end{equation}
where \emph{N} = number of detected ions, $T_{ex}$ = excitation time and \emph{K} is an empirical constant independent of ion number and excitation 
time. In this experiment the precision was maximized by collecting a large number of ions and by applying long excitation times. Also a higher 
$\nu_{c}$ improves the relative precision achieved with a strong magnetic field (see Eq.~(1)). In addition, the Ramsey excitation yields a narrower 
central peak compared to the conventional excitation scheme. Therefore, using the Ramsey excitation the precision is improved by a factor of about 2 
or more \cite{SG07a}. All these factors together allow us to determine the \emph{Q}-value of the double-beta decay of $^{100}$Mo with high-precision.

\begin{figure}
\resizebox{0.9\textwidth}{!}{
\includegraphics{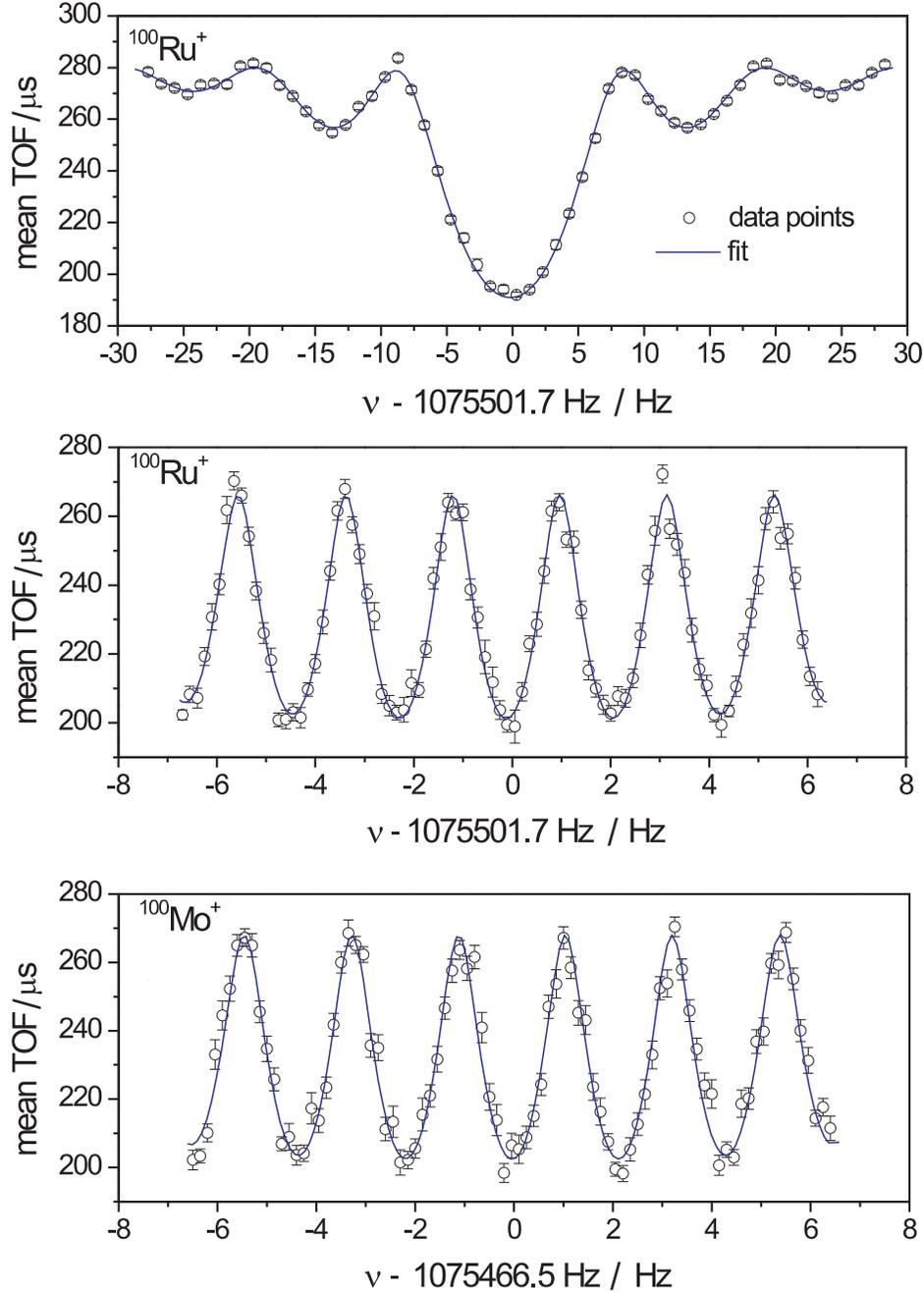}}
\caption{\label{fig:singlecolumn} Time-of-flight (TOF) resonances of $^{100}$Ru and $^{100}$Mo ions from the precision trap. The top panel represents 
a typical TOF resonance using the conventional excitation scheme with an excitation time of 100 ms. The middle and bottom panels display the TOF 
resonances for $^{100}$Ru and $^{100}$Mo ions using the Ramsey excitation pattern with 50-400-50 ms, respectively.}
\end{figure}

For the consistency of the setup we have measured the \emph{Q}-value of the double-beta decay of $^{76}$Ge which was measured in 2001 at SMILETRAP 
\cite{GD01} and recently again in 2007 applying the Ramsey excitation method \cite{MS07}. Our value agrees with the SMILETRAP value (2039.006(50) keV) 
\cite{GD01} within the error bars. For $^{76}$Ge 25 resonances were collected with $^{76}$Se as a reference. A Ramsey excitation scheme was applied 
with two time-separated fringes of 25 ms with a waiting time of 250 ms (25-250-25 ms). For $^{100}$Mo a total of 68 resonances were collected with 
$^{100}$Ru as a reference. These were grouped in two sets with Ramsey excitation times of  25-250-25 ms and 50-400-50 ms. Figure 3 shows the 
scattering of the individual frequency ratios $r_{i}$ relative to the weighted average value of the frequency ratio $\bar{r}$ =  1.000 032 6069(18). 
The inner and outer statistical uncertainties ($\delta \bar{r}$) \cite{RT32} of the weighted average frequency ratio are 1.6$ \times 10^{-9}$  and 1.8 
$\times 10^{-9}$, respectively. The ratio of these values (Birge ratio) is very close to 1 which confirms that the scattering of the data is 
statistical. However, the larger error (in this case outer error) was used to determine the final uncertainty.          

\begin{figure}
\resizebox{0.9\textwidth}{!}{
\includegraphics{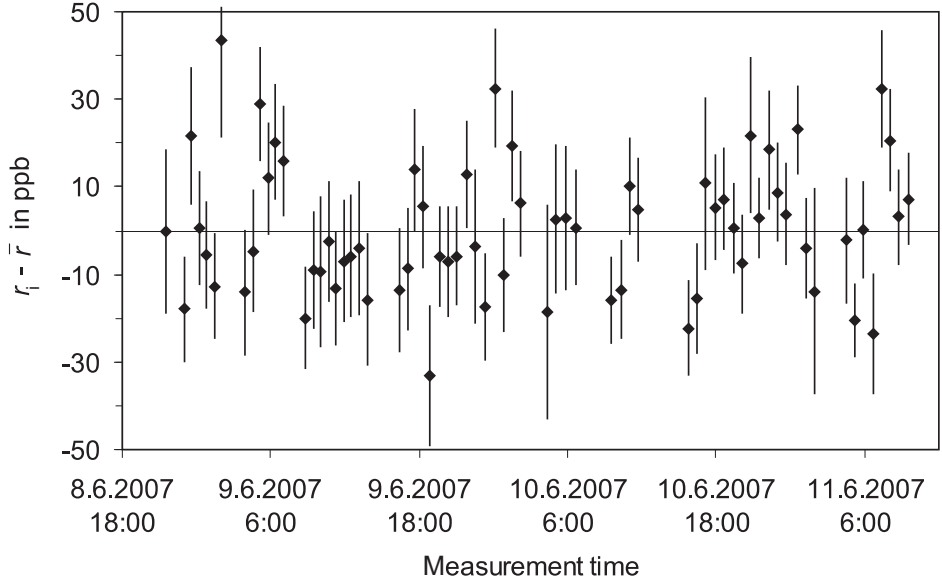}}
\caption{\label{fig:epsart}Measured cyclotron frequency ratios $r_{i}$ between $^{100}$Mo$^{+}$ and $^{100}$Ru$^{+}$ relative to the weighted average 
value $\bar{r}$.}
\end{figure}

In the determination of the cyclotron frequency ratio the following systematic uncertainties were taken into account: The number of ions present in 
the trap can cause a shift in the cyclotron frequency. This is taken into account by plotting the center value of the cyclotron resonance as a 
function of the detected number of ions \cite{AK03}. The cyclotron frequency equivalent to one ion in the trap was determined from this plot via a 
linear extrapolation to the observed 0.6 ions (detector efficiency = 60\%). This extrapolated cyclotron frequency was taken as a final value. Hence 
the uncertainty due to the count rate comes together with the statistical uncertainty. 

The drift of the magnetic field was taken into account by an interpolation of the reference frequencies measured before and after the cyclotron 
frequency measurement of the ion of interest. This linear interpolation does not take into account the short term fluctuations of the magnetic field. 
This was taken into account by adding $\Delta \mbox{B}/\mbox{B}$ = 3.22(16)$\times 10^{-11}/$min \cite{SR07a} multiplied by the time difference 
between the two consecutive reference measurements quadratically to the uncertainty of the frequency ratio. As the \emph{Q}-value is determined by the 
cyclotron frequency ratio between the mother and daughter having same $\frac{A}{q}$, mass-dependent and other systematic uncertainties (if it exists) 
should cancel. 

The \emph{Q}-value can be derived from the final weighted average frequency ratio using Eq.~(4). The term inside the parenthesis in Eq.~(4) is small 
($\sim 2-3\times10^{-5}$), thus the uncertainty contribution from $m_{d}$ to the \emph{Q}-value is negligible. Figure 4 shows the weighted average 
Q-values of $^{100}$Mo with their uncertainties for two different excitation times and these two sets of data agree each other. The final weighted 
average cyclotron frequency ratios $\bar{r}$ and Q-values of $^{76}$Ge and $^{100}$Mo with their corresponding uncertainties in the parenthesis are 
given in Table 1.

\begin{table}
\caption{\label{tab:table3}Weighted average frequency ratios ($\bar{r}$) and the \emph{Q}-values of $^{76}$Ge and $^{100}$Mo measured at JYFLTRAP. The 
final uncertainty normalized with the $\chi^{2}$ is given in the parenthesis. \# and $T_{ex}$ represent the number of doublet measurements and 
excitation scheme, respectively. Final frequency ratios and \emph{Q}-values are indicated in bold.}
\begin{center}
\begin{tabular}{llccll} \hline
 Mother&Daughter& \# & $T_{ex}$ (ms)& Frequency ratio, $\bar{r}$ = $\frac{\nu_{d}}{\nu_{m}}$& \emph{Q}-value (keV) \\ \hline
 $^{76}$Ge & $^{76}$Se &21 &25-250-25&\textbf{1.000 028 8332(21)}& \textbf{2039.04(16)} \\ \hline 
 $^{100}$Mo&$^{100}$Ru&22& 25-250-25&1.000 032 6056(33)&\\
 $^{100}$Mo&$^{100}$Ru&46& 50-400-50&1.000 032 6074(20)&\\
 $^{100}$Mo  & $^{100}$Ru & 68&  & \textbf{1.000 032 6069(18)} &\textbf{3034.40(17)}\\\hline
\end{tabular}
\end{center}
\end{table}

\begin{figure}
\resizebox{0.90\textwidth}{!}{
\includegraphics{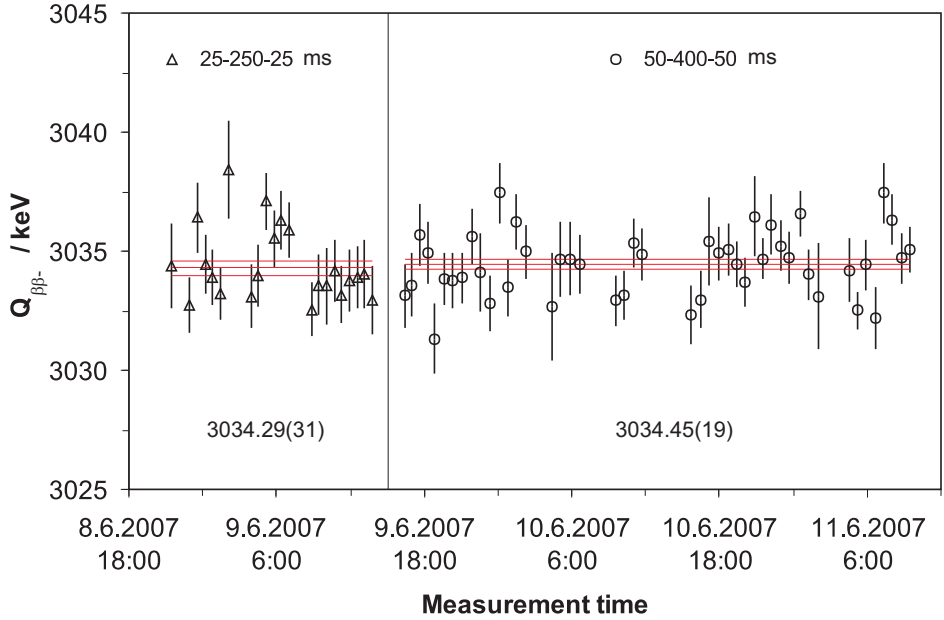}}
\caption{\label{fig:epsart}\emph{Q}-values obtained from individual measurements of $^{100}$Mo in
June 2006 by using  $^{100}$Ru as a reference. Triangles and circles correspond to different excitation times (see
legend). The data collection was stopped every night before the automatic $B_{0}$ dump at 3:00 AM. The statistical, count-rate and magnetic-field 
fluctuation uncertainties have been taken into account for each set.}
\end{figure}

\section{Discussion}
The precise $^{100}$Mo \emph{Q}-value is used to calculate the phase-space integrals \emph{G} of the $2\nu\beta\beta$-decay and 
$0\nu\beta\beta$-decays. A detailed formula for these calculations can be found in ref. \cite{JS98}. Results of the calculations are summarized in 
Table~2. The uncertainty in \emph{G} is estimated solely from our \emph{Q}-value uncertainty. The other uncertainty (in addition to the uncertainty 
from the \emph{Q}-value) in \emph{G} comes from the Fermi function approximation for the exact solution of the Dirac equation for a homogeneously 
charged sphere. The relativistic Fermi approximation limits the accuracy of \emph{G} to some three digits.
  
\begin{table}
\caption{\label{tab:table3}Phase-space integrals for $^{100}$Mo. The axial-vector coupling constant g$_{A}$ = 1.0 and 1.254 and electron rest mass 
$m_{e}$ =  510.998918 keV were used in the calculations. The uncertainty for \emph{G} is estimated solely from the \emph{Q}-value uncertainty.}
\begin{center}
\begin{tabular}{lllll}\hline
 Q-value / keV & \vline  $G^{2\nu} \times 10^{-18}$ &for &\vline $G^{0\nu} \times 10^{-14}$&for\\
   &\vline $g_{A}=1.0$& $g_{A}=1.254$ &\vline  $g_{A}=1.0$& $g_{A}=1.254$\\\hline
 3034.40(17)&\vline 3.8179(19) &9.4409(47)  &\vline 1.8915(4)& 4.6772(10) \\ \hline
 3035(6)    &\vline 3.826(67) &9.46(17)&\vline1.893(14) & 4.681(34)\\\hline
\end{tabular}
\end{center}
\end{table}

\begin{table}
\caption{\label{tab:table3}A comparison between the calculated and experimental (Ex) nuclear matrix elements M$^{2\nu}$ for $^{100}$Mo. For detailed 
notation see ref. \cite{JS98}.}
\begin{center}
\begin{tabular}{lllllll}\hline
 Ex &Ex &QPRA &QPRA &SPRA &SU(3) & SU(3)\\ for&for &EMP &EMP &WS &SPH &DEF \\ $g_{A}$=1.0&$g_{A}$=1.254& \cite{AG92}& \cite{JS94}&\cite{SS95}& 
\cite{JG95}& \cite{JG95}\\
 \hline
 0.192(4) &0.122(4)& 0.256& 0.197&0.059 &0.152 &0.108 \\\hline
\end{tabular}
\end{center}
\end{table}

Using Eq.~(1) the value of the nuclear matrix element $M^{2\nu}$ is estimated experimentally  to be 0.122$\pm0.004$ where the precise phase-space 
integral value is taken from this work and the recommended half-life $ T_{1/2} = (7.1 \pm 0.4) \times 10^{18}$ years is taken from \cite{ASB06}. A 
comparison between the computed nuclear matrix elements and the experimental one is shown in Table~3. Theoretical values of $M^{2\nu}$ vary by a 
factor of five whereas the experimental value lies in between. The matrix elements 0.197 and 0.152 of the spherical QRPA and SU(3) theories are 
consistent with the range of the experimental matrix elements. On the other hand the matrix element 0.108 computed by using the deformed SU(3) theory 
falls out side this range. This would suggest that deformation is not very important for the ground-state to ground-state $2\nu\beta\beta$ decay of 
$^{100}$Mo.


In the case of $0\nu\beta\beta$-decay finding an estimate for the experimental value of the nuclear matrix element $M^{0\nu}$ is not so 
straight-forward since the decay half-life is unknown in Eq.~(2). However the expected half-life can be plotted as a function of the effective 
neutrino mass as shown in Fig.~5. We have used our value for the phase-space integral and a set of nuclear matrix elements $M^{0\nu}$ calculated by 
using the pnQRPA \cite{MK07,FS07} and RQRPA \cite{FS07} models. In these models the nucleon-nucleon short-range correlations have been accounted for 
by two different methods, the Unitary Correlation Operator Method (UCOM) and Jastrow. For both models, the pnQRPA and the RQRPA, the range of computed 
matrix elements stems from the uncertainty in the value of the axial-vector coupling coefficient, $g_{A}$ = 1.0 - 1.25, and from the variation in the 
value of the proton-neutron particle-particle interaction strength $g_{pp}$, used to fit the experimental range of the  $2\nu\beta\beta$ half-life. 

\begin{figure}
\resizebox{0.90\textwidth}{!}{
\includegraphics{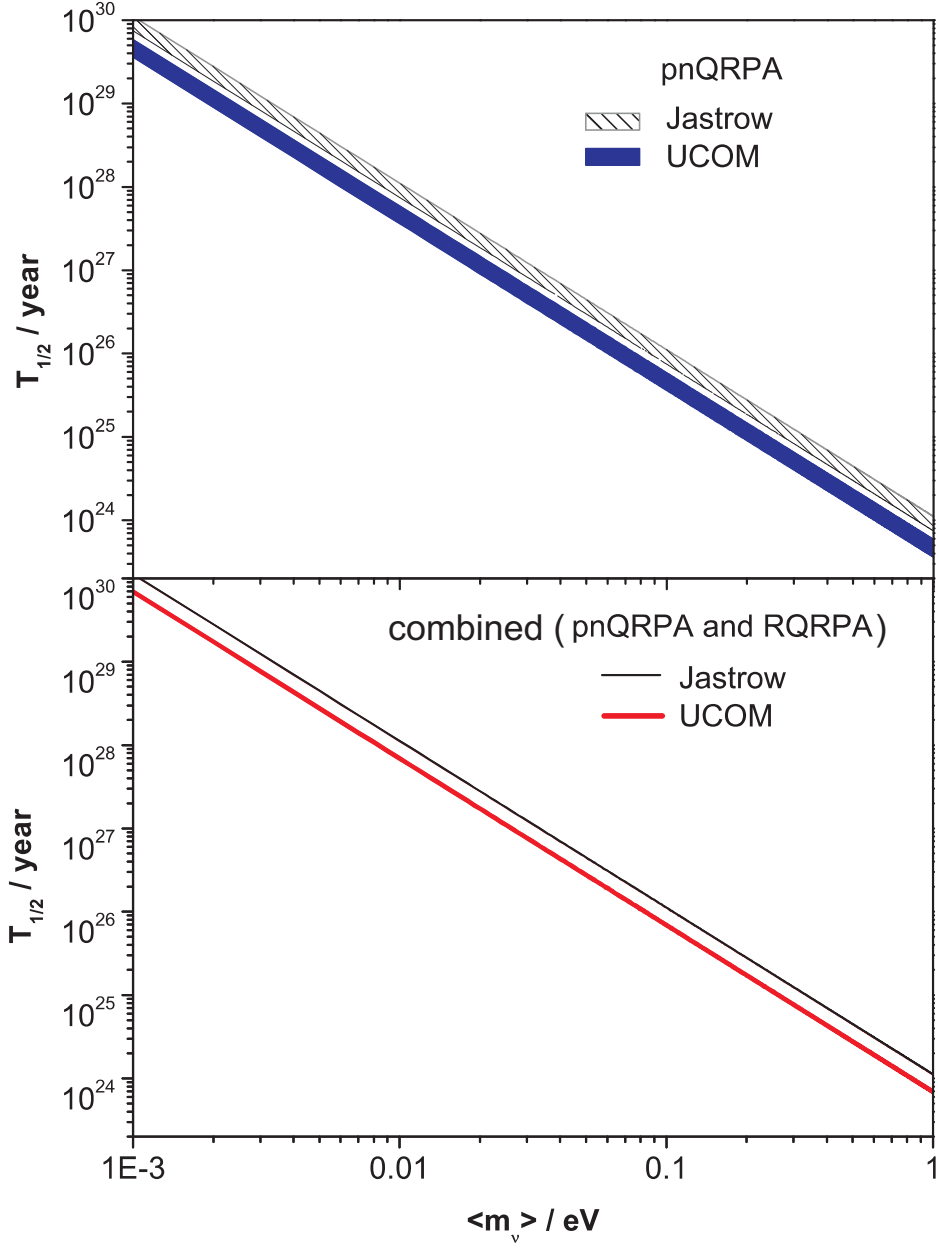}}
\caption{\label{fig:epsart}Expected half-life band as a function of the effective neutrino mass for $^{100}$Mo. The ranges of the nuclear matrix 
element M$^{0\nu}$ come from \cite{MK07} in the upper panel and from \cite{FS07} in the lower panel. The values of the phase-space integrals were used 
from this work.}
\end{figure}

 \section{Conclusion}
In this article an off-line ion source is demonstrated for creating stable (or long-lived) mixed beams of single-charged ions. A precise double-beta 
decay \emph{Q}-value, 3034.40$\pm 0.17$ keV for $^{100}$Mo, is determined, a candidate for searching the evidence of $0\nu\beta\beta$-decay. The 
\emph{Q}-value precision of about 200 eV was obtained for mass 100 region. Experimental phase-space integral \emph{G} is obtained and using this value 
an experimental $M^{2\nu}$ value is derived for $^{100}$Mo. In the case of neutrinoless double-beta decay half-life is estimated as a function of the 
effective neutrino mass for a set of nuclear matrix elements.

\begin{ack}                               
This work has been supported by the TRAPSPEC Joint Research Activity project under the EU 6th Framework program "Integrating Infrastructure Initiative 
- Transnational Access", Contract Number: 506065 (EURONS) and by the Academy of Finland under the Finnish Center of Excellence Program 2006-2011 
(Nuclear and Accelerator Based Physics Program at JYFL and project number 202256 and 111428)..  
\end{ack}


\begin{thebibliography}{99}     
\bibitem{sk}
Y. Fukuda \emph{et al.}, (Super-Kamiokande Collaboration), Phys.
Rev. Lett. \textbf{81}, 1562 (1998).
\bibitem{sn}
Q. R. Ahmad \emph{et al.}, (SNO Collaboration), Phys. Rev. Lett.
\textbf{87}, 071301 (2001).
\bibitem{kl}
K. Eguchi \emph{et al.}, (KamLAND Collaboration), Phys. Rev.
Lett. \textbf{90}, 021802 (2003).
\bibitem{LB99}
L. Baudis \emph{et al.}, Phys. Rev. Lett. \textbf{83}, 41 (1999).

\bibitem{HV011}
H.V. Klapdor-Kleingrothaus \emph{et al.}, Mod. Phys. Lett. A \textbf{16}, 2409 (2001).
\bibitem{CE02}
C.E. Aalseth \emph{et al.}, Mod. Phys. Lett. A \textbf{17}, 1475 (2002).
\bibitem{HV01}
 H.V. Klapdor-Kleingrothaus \emph{et al.}, Eur. Phys. J. A \textbf{12}, 147 (2001).
 \bibitem{HE91}H. Ejiri \emph{et al.}, Phys. Lett. B \textbf{258}, 17 (1991).
\bibitem{MN05} M. Nomachi \emph{et al.}, Nucl. Phys. B \textbf{138}, 221 (2005). 
\bibitem{RA05} R. Arnold \emph{et al.}, Nucl. Inst. Meth. A \textbf{536}, 79 (2005). 
\bibitem{JS98} J. Suhonen and O. Civitarese, Phys. Rep. \textbf{300}, 123 (1998). 
\bibitem{AJ06} A. Jokinen \emph{et al.}, Int. J. Mass Spect. \textbf{251}, 204 (2006).
\bibitem{UH06} U. Hager \emph{et al.}, Phys. Rev. Lett \textbf{96}, 042504 (2006).
\bibitem{AK06} A. Kankainen \emph{et al.}, Eur. Phys. J. A \textbf{29}, 271 (2006).
\bibitem{SR07} S. Rahaman \emph{et al.}, Eur. Phys. J A \textbf{32}, 87 (2007).
\bibitem{SR07a} S. Rahaman \emph{et al.}, Eur. Phys. J A \textbf{34}, 5 (2007).
\bibitem{UH07} U. Hager \emph{et al.}, Phys. Rev. C \textbf{75}, 064302 (2007).
\bibitem{TE06} T. Eronen \emph{et al.}, Phys. Lett. B \textbf{636}, 191 (2006).
\bibitem{TE06a} T. Eronen \emph{et al.}, Phys. Rev. Lett. \textbf{97},  232501 (2006).
\bibitem{CW07} A. Jokinen\emph{ et al.}, Hyper. Interact., \textbf{173}, 143 (2006).
\bibitem{GB86} L.S. Brown and G. Gabrielse, Rev. Mod. Phys. \textbf{58}, 233 (1986). 
\bibitem{JA01} J. Äystö, Nucl. Phys. A \textbf{693}, 477 (2001).
\bibitem{GA03} G. Audi, A.H. Wapstra and C. Thibault, Nucl. Phys. A \textbf{729}, 337 (2003).
\bibitem{AN03} A. Nieminen \emph{et al.}, Nucl. Instrum. Meth. Phys. Res. B \textbf{204}, 563 (2003).
\bibitem{GS91} G. Savard \emph{et al.}, Phys. Lett. A \textbf{158}, 247 (1991). 
\bibitem{NFR90} N.F. Ramsey, Rev. Mod. Phys. \textbf{62}, 541 (1990).
\bibitem{TE07} T. Eronen \emph{et al.}, submitted to EMIS Conf. Proc. (2007).
\bibitem{TE08} T. Eronen \emph{et al.}, submitted to Phys. Rev. Lett. (2007).
\bibitem{MK95} M. König \emph{et al.}, Int. J. Mass Spectr. Ion Proc. \textbf{142}, 95 (1995).
\bibitem{SG07} S. George \emph{et al.}, Phys. Rev. Lett. \textbf{98}, 162501 (2007).
\bibitem{GB01} G. Bollen, Nucl. Phys. A \textbf{693}, 3 (2001).
\bibitem{SG07a} S. George \emph{et al.}, Int. J. Mass Spect. \textbf{264}, 110 (2007).
\bibitem{GD01} G. Douysset, T. Fritioff and C. Carlberg, Phys. Rev. Lett. \textbf{86}, 4259 (2001).
\bibitem{MS07}M. Suhonen \emph{et al.}, J. of Instrum. \textbf{2}, 06003 (2007). 
\bibitem{RT32} R.T. Birge, Phys. Rev. \textbf{40}, 207 (1932). 
\bibitem{AK03}A. Kellerbauer \emph{et al.},  Eur. Phys. J. D \textbf{22}, 53 (2003). 
\bibitem{ASB06}
A.S. Barabash, Czechoslovak J. of Phys. 
\textbf{56}, 437 (2006).
\bibitem{AG92}
A. Griffiths and P. Vogel, Phys. Rev. C \textbf{46}, 181 (1992).
\bibitem{JS94}
J. Suhonen and O. Civitarese, Phys. Rev. C \textbf{49}, 3055 (1994).
\bibitem{SS95}
S. Stoica, Phys. Lett B \textbf{350}, 152 (1995).
\bibitem{JG95}
J.G. Hirsch \emph{et al.}, Phys. Rev. C \textbf{51}, 2252 (1995).
\bibitem{MK07}
M. Kortelainen and J. Suhonen, Phys. Rev. C \textbf{76}, 024350 (2007).
\bibitem{FS07}
F. \v Simkovic \emph{et al}., arXiv:0710.2055v2 [nucl-th] (2007).

\end{thebibliography}

\end{document}